\documentclass[prb,twocolumn,showpacs,superscriptaddress,preprintnumbers,amsmath,amssymb]{revtex4-1}
\usepackage{graphicx}
\usepackage{dcolumn}
\usepackage{bm}
\usepackage{amsmath}
\usepackage{color}
\usepackage{hyperref}
\usepackage[utf8]{inputenc}
\usepackage[T1]{fontenc}

\begin{document}


\title{Dynamical Dzyaloshinsky-Moriya interaction in KCuF$_3$: Raman evidence for an antiferrodistortive lattice instability}

\author{V.~Gnezdilov}
\affiliation{B.I. Verkin Inst. for Low Temperature Physics and
Engineering, NASU, 61103 Kharkov, Ukraine}

\author{J.~Deisenhofer}
\affiliation{Experimental Physics~V, Center for Electronic
Correlations and Magnetism, University of Augsburg,
D-86135~Augsburg, Germany}

\author{P.~Lemmens}
\author{D.~Wulferding}
\affiliation{Institute for Condensed Matter Physics, Braunschweig University of Technology, D-38106~Braunschweig, Germany}

\author{O. Afanasiev}
\affiliation{B.I. Verkin Inst. for Low Temperature Physics and
Engineering, NASU, 61103 Kharkov, Ukraine}

\author{P. Ghigna}
\affiliation{Dipartimento di Chimica Fisica, Universit$\grave{a}$ di Pavia,
I-27100 Pavia, Italy}

\author{A.~Loidl}
\affiliation{Experimental Physics~V, Center for Electronic
Correlations and Magnetism, University of Augsburg,
D-86135~Augsburg, Germany}

\author{A. Yeremenko}
\affiliation{B.I. Verkin Inst. for Low Temperature Physics and
Engineering, NASU, 61103 Kharkov, Ukraine}

\date{\today}

%

\begin{abstract}
In the orbitally ordered, quasi-one  dimensional Heisenberg
antiferromagnet KCuF$_3$ the low-energy $E_g$ and $B_{1g}$ phonon
modes show an anomalous softening ($\sim$25$\%$ and $\sim$13$\%$)
between room temperature and the characteristic temperature $T_S$ =
50 K. In this temperature range a freezing-in of F ion dynamic
displacements is proposed to occur. In addition, the $E_g$ mode at
about 260~cm$^{-1}$ clearly splits below $T_S$. The width of the
phonon lines above $T_S$ follows an activated behavior with an
activation energy of about 50~K. Our observations clearly evidence a
reduction of the structural symmetry below $T_S$ and indicate a
strong coupling of lattice and spin fluctuations for $T>T_S$.

\end{abstract}


\maketitle

\section{\label{sec:level1}Introduction}

The system KCuF$_3$ has long been known as a paradigm for an
orbitally ordered system where a cooperative Jahn-Teller (JT)
distortion is strongly competing with the electronic degrees of
freedom as the driving force behind the orbital
order.\cite{kadota,kugel,liechtenstein,medvedeva} This system was
investigated recently by realistic band structure calculations as a
benchmark system for modeling structural relaxation effects due to
electronic correlations~\cite{leonov,leonov09} and for revealing the
influence of electronic superexchange on the orbital
ordering.\cite{pavarini} The compound seems to be orbitally ordered
throughout its solid phase, but shows long-range A-type
antiferromagnetic (AFM) ordering only below $T_N$ = 39 K. In
literature an orbital ordering temperature of about 800~K is often
evoked in this system, but astonishingly experimental evidence for a
transition at this temperature seems to be evasive. Early on,
however, it was reported that between 670~K and 720~K an
irreversible transition takes place.\cite{okazaki61} Recently, the
melting of the cooperative JT-transition has been studied in
KCu$_{1-x}$Mg$_{x}$F$_3$ and from the extrapolation to undoped
KCuF$_3$ a JT transition temperature of 1350~K has been
estimated.\cite{Ghigna10} The paramagnetic (PM) susceptibility has
been described by a Bonner-Fisher law with an exchange constant J =
190 K,\cite{kadota} indicating that the compound is a good
realization of a one-dimensional (1D) spin chain in the PM regime.
Inelastic neutron scattering studies did reveal a spinon-excitation
continuum, a clearly 1D quantum phenomenon, existing also below the
N\'eel temperature.\cite{lake1,lake2} From a structural point of
view the reported relatively high tetragonal
symmetry~\cite{okazaki,buttner,tsukuda,hutching1,hutching2,satija}
($D^{18}_{4h}$ -- $I$4$/mcm$) makes KCuF$_3$ one of the simplest
systems to study. However, the established symmetry assignment has
been questioned by an X-ray diffraction investigation~\cite{hidaka}
which suggested the existence of orthorhombic distortions in
KCuF$_3$ at room temperature with $D_2^4$ symmetry. A
low-temperature Raman scattering study~\cite{ueda} revealed a
difference of spectra measured in $xz$ and $yz$ polarization and
anomalously broad linewidths of the stretching modes, which was
interpreted as evidence of a symmetry lower than $D^{18}_{4h}$ also
below the N\'eel temperature. Although orthorhombic distortions were
involved for explaining the electron spin resonance (ESR) properties
of KCuF$_3$,\cite{yamada1} discrepancies remain for the analysis of
recent NQR,\cite{mazzoli} AFM resonance,\cite{li} and further
experimental and theoretical findings.\cite{yamada2,binggeli}
Besides, in X-ray resonant scattering\cite{paolasini,caciuffo} of
the orbital ordering (OO) in KCuF$_3$ indications for a coupling of
lattice and magnetic degrees of freedom above $T_N$ were found. Only
recently, the ESR properties for $T > T_N$ could be successfully
explained within the tetragonal symmetry by assuming a dynamical
Dzyaloshinsky-Moriya (DM) interaction related to strong oscillations
of the bridging F$^-$ ions perpendicular to the crystallographic $c$
axis.\cite{eremin} It was argued that these dynamic distortions
freeze in at a temperature $T_S$ = 50 K, leading to an effectively
lower symmetry and the occurrence of exciton-magnon sidebands in
optical absorption experiments.\cite{deisenhofer}

Here we report on a detailed study of the temperature dependence of the Raman-active phonons in a KCuF$_3$ single crystal tracking the
symmetry reduction during the anticipated freezing of the dynamic distortion at $T_S$ = 50 K and the N\'eel ordering at $T_N$ = 39 K. We
find a large softening of the lowest lying $E_g$ mode and the $B_{1g}$ mode by 25\% and 13\% between room temperature and $T_S$, respectively. The linewidth and the integrated intensity of these modes also exhibit anomalies at $T_S$ and $T_N.$  Moreover, the $E_g$ mode at about 260~cm$^{-1}$ clearly splits below $T_S$ evidencing the existence of an antiferrodistortive lattice instability in KCuF$_3$ which leads
to a symmetry reduction at $T_S$ = 50 K prior to magnetic ordering.

\section{\label{sec:level1}Experimental Details}

The single crystal was oriented by Laue diffraction, cut along the
(110) pseudocubic plane and mechanically polished to optical
quality. Details on crystal growth are described in
Ref.~\onlinecite{caciuffo}. The sample has previously been
investigated by ESR and optical
spectroscopy.\cite{eremin,deisenhofer} The Raman spectra were
obtained with two different experimental setups and in two
geometries of experiment: (i) a DILOR XY triple spectrometer with a
liquid-nitrogen-cooled CCD detector (quasi-backscattering geometry)
and (ii) a U1000 high resolution double spectrometer with RCA 31034A
photomultiplier (right-angle scattering geometry). The 647 nm Ar/Kr
ion (5 mW output power) and the 632.8 nm He-Ne (25 mW output power)
lasers were used for excitation in these two setups, respectively.
Temperature dependencies were obtained in variable temperature
gas-flow cryostats.

\section{\label{sec:level1}Experimental Results and Discussion}

\begin{figure}
 \centering
\includegraphics[width=8cm]{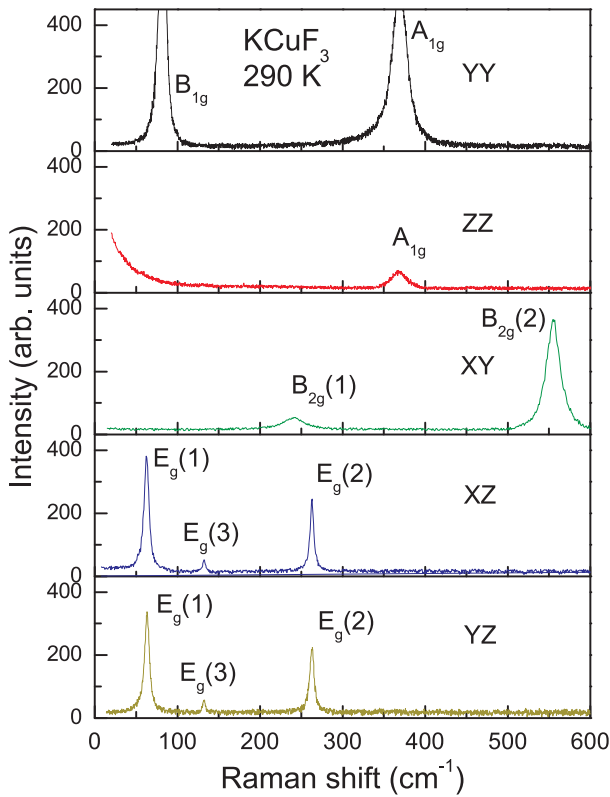}
\caption{\label{fig:fig1} Polarized Raman spectra of single crystal KCuF$_3$ taken at
290 K in different scattering configurations.}
\end{figure}
\begin{figure}
 \centering
\includegraphics[width=8cm]{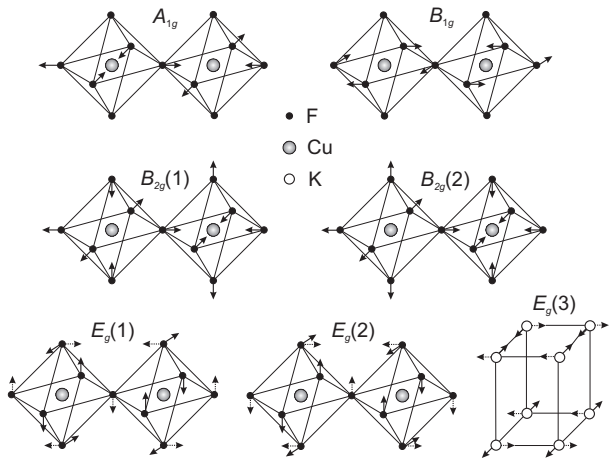}
\caption{\label{fig:fig2} Raman allowed phonon modes for the KCuF$_3$ with
$D^{18}_{4h}$ structure. The degeneracy of the $E_g$ modes is
indicated using solid and dotted arrows.}
\end{figure}
\begin{figure}[b]
 \centering
\includegraphics[width=8cm]{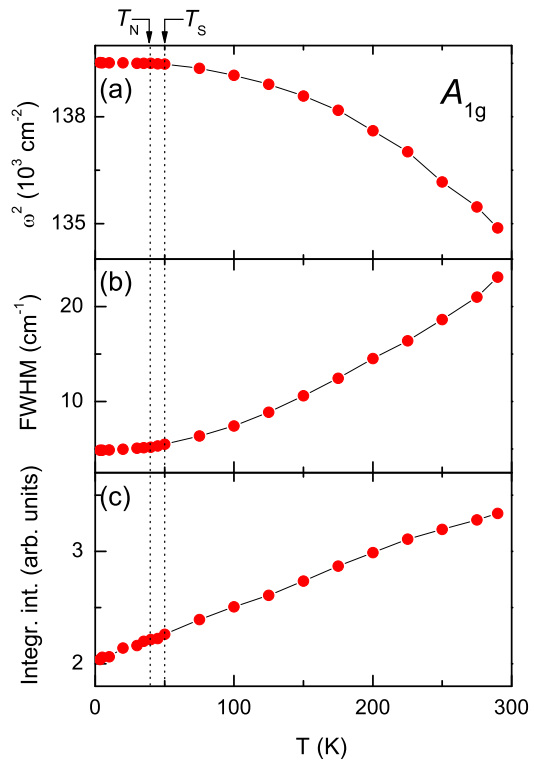}
\caption{\label{fig:A1goverview} Parameters of the $A_{1g}$ mode:
Temperature dependence of (a) the squared eigenfrequency $\omega_0$,
(b) the FWHM linewidth, and (c) the Bose corrected integrated
intensity. Lines are to guide the eye.}
\end{figure}

In Fig.~\ref{fig:fig1} the polarized Raman spectra of single
crystalline KCuF$_3$ taken in $yy$, $zz$, $xy$, $xz$, and $yz$
scattering configurations are shown for $T$ = 290 K. The number of
lines and the selection rules are fully consistent with the
theoretically expected Raman-active normal modes\cite{ueda} of
KCuF$_3$ with tetragonal $D^{18}_{4h}$
\begin{equation}
\Gamma_{\rm Ram} = A_{1g}(yy,zz) + B_{1g}(yy) + 2B_{2g}(xy)
+3E_g(xz,yz)
\end{equation}
Hence, the three lines in both the $xz$ and $yz$ spectra correspond
to the three $E_g$ modes. The line observed with different
intensities in $yy$ and $zz$ spectra is identified as the $A_{1g}$
mode. The intense line observed only in the $yy$ spectrum can be
assigned to the $B_{1g}$ mode. Finally, the two lines in the $xy$
spectra are the two $B_{2g}$ modes. At room temperature all lines
have a Lorentzian lineshape. Figure~\ref{fig:fig2} shows
schematically  the vibrational patterns for the seven Raman-active
modes of each symmetry ($A_{1g}$, $B_{1g}$, $B_{2g}$, and $E_g$) of
KCuF$_3$ derived from the $D^{18}_{4h}$ space group. The observed
spectra and mode assignments are in agreement with previously
reported data at 10~K.\cite{ueda} A direct comparison of our data at
4~K and 290~K with Ref.\onlinecite{ueda} and theoretical
estimates\cite{nikiforov96} is presented in Tab.~\ref{fig:tab1}. In
general, there is a good agreement between the corresponding values
except for the $B_{2g}$(1) mode with a frequency of 240.4 cm$^{-1}$
observed in our experiments in contrast to a somewhat higher
frequency of 265.8 cm$^{-1}$ in Ref.~\onlinecite{ueda}. The second
discrepancy is that the lines assigned to $E_g$(1,2) and $B_{1g}$
are almost two times broader in the low-temperature Raman spectra of
Ref.~\onlinecite{ueda}. The phonon lines of $A_{1g}$ and $B_{2g}$
symmetry have large linewidths in comparison with the other modes.
In Fig.~\ref{fig:A1goverview} we show the temperature dependent
parameters for the $A_{1g}$ mode as an example. The $A_{1g}$ and
$B_{2g}$ modes, aside from their broadened lineshape, show no
anomalous behavior. In the full temperature range they exhibit a
hardening of 1-2$\%$.

Moreover, we observe quasielastic scattering in $zz$ configuration,
which is a general feature in low-dimensional spin
systems\cite{lemmens} and should only be observed in intra-chain
scattering configuration, i.e. with the light polarization parallel
to the effective chain direction. This quasielastic scattering in
KCuF$_3$ has been investigated in detail by Yamada and Onda
previously\cite{yamada3} and will not be further considered in our
work.

\begin{table*}[t]
\caption{\label{fig:tab1} Frequencies and linewidths of the observed
Raman modes in cm$^{-1}$ in KCuF$_3$ at 4~K and 290~K compared to
the experimental values reported in Ref.\onlinecite{ueda} at 10~K
and calculations from Ref.\onlinecite{nikiforov96}.}
\begin{ruledtabular}
\begin{tabular}{c@{\hspace{3em}}*{7}{c}}
    Mode      & \multicolumn{4}{c}{ Frequency (cm$^{-1}$)}       &\multicolumn{3}{c} {Linewidth (cm$^{-1}$)} \vspace{1mm}  \\
              & 290K   & 4 K  & 10 K (Ref.\onlinecite{ueda})   &Calculated (Ref.\onlinecite{nikiforov96}) & 290K      & 4K & 10 K (Ref.\onlinecite{ueda})
              \vspace{0.5mm}\\\hline
  $A_{1g}$    & 367.3    & 373.5    & 374.8  & 398   & 23.1      & 4.9   & 9.2 \\
  $B_{1g}$    & 81.6     & 70.9     & 72.8   & 100    & 7.1       & 0.9   & 1.6 \\
  $B_{2g}$(1) & 240.4    & 245.2    & 265.8  & 259   & 30.8      & 8.7   & 7.0 \\
  $B_{2g}$(2) & 554.8    & 561.3    & 563.0  & 586   & 22.6      & 9.1   & 9.1 \\
  $E_{g}$(1)  & 63.0     & 47.4     & 53.2   & 50   & 5.8       & 0.7   & 3.0 \\
  $E_{g}$(2)  & 262.9    & 260.8    & 261.6  & 136   & 7.5       & 1.7   & 3.0 \\
  $E_{g}$(3)  & 132.3    & 129.3    & 131.2  & 268    &7.5       & 1.6   & 1.6 \\

\end{tabular}
\end{ruledtabular}
\end{table*}

In the following we will focus on the temperature dependence of the
modes $E_g(1)$, $E_g(2)$  and $B_{1g}$. The $E_g$(1) mode reportedly
exhibits a weak splitting at 10~K only when measured in
$yz$-configuration. In contrast, the $E_g$(2) mode shows a splitting
only when measuring in $xz$-configuration. The $E_g$(3) mode, which
corresponds to a vibration of K$^+$ ions, shows no splitting in
either of the two configurations.\cite{ueda} The $E_g$(1) and
$E_g$(2) modes correspond to shearing vibrations of the F$^-$ ions
which involves a displacement of the fluorine ions away from the
Cu-F-Cu bonding lines, while the $B_{1g}$ mode corresponds to a
tilting motion of the F$^-$ ions around the central Cu atom (see
Fig.~\ref{fig:fig2}). As such displacements are thought to be the
origin of the dynamical DM interaction which allows to understand
the ESR and antiferromagnetic resonance properties, we expect that
these modes are strongly related to the proposed freezing of the
dynamic fluorine displacements below $T_S$ =
50~K.\cite{eremin,deisenhofer}

\begin{figure}[b]
 \centering
\includegraphics[width=8cm]{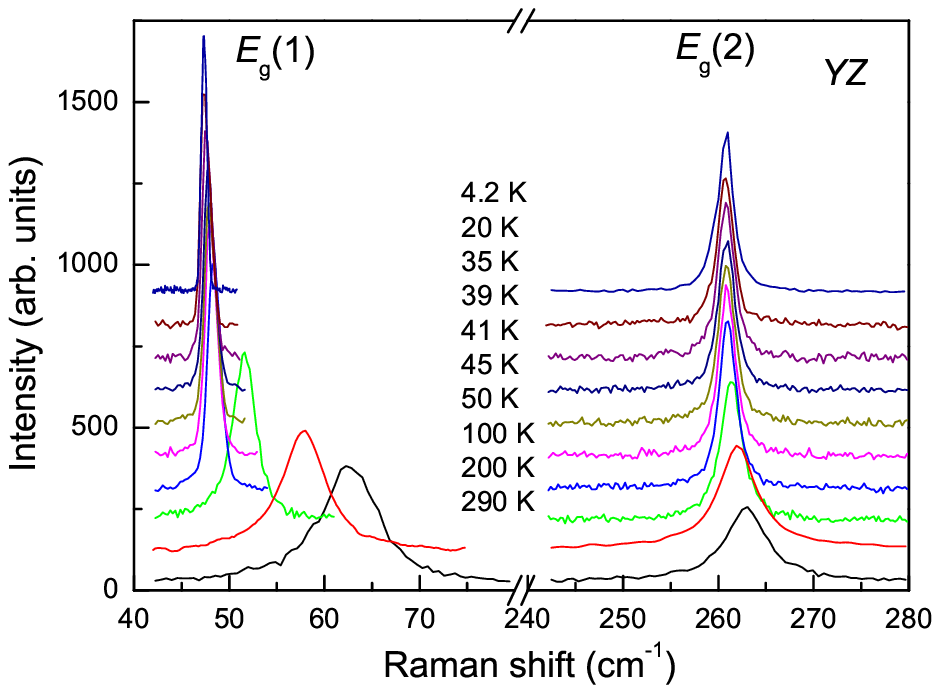}
\caption{\label{fig:fig3} Temperature dependent Raman spectra of the $E_g$(1) and $E_g$(2) modes taken in the $yz$ scattering geometry.}
\end{figure}

\begin{figure}
 \centering
\includegraphics[width=8cm]{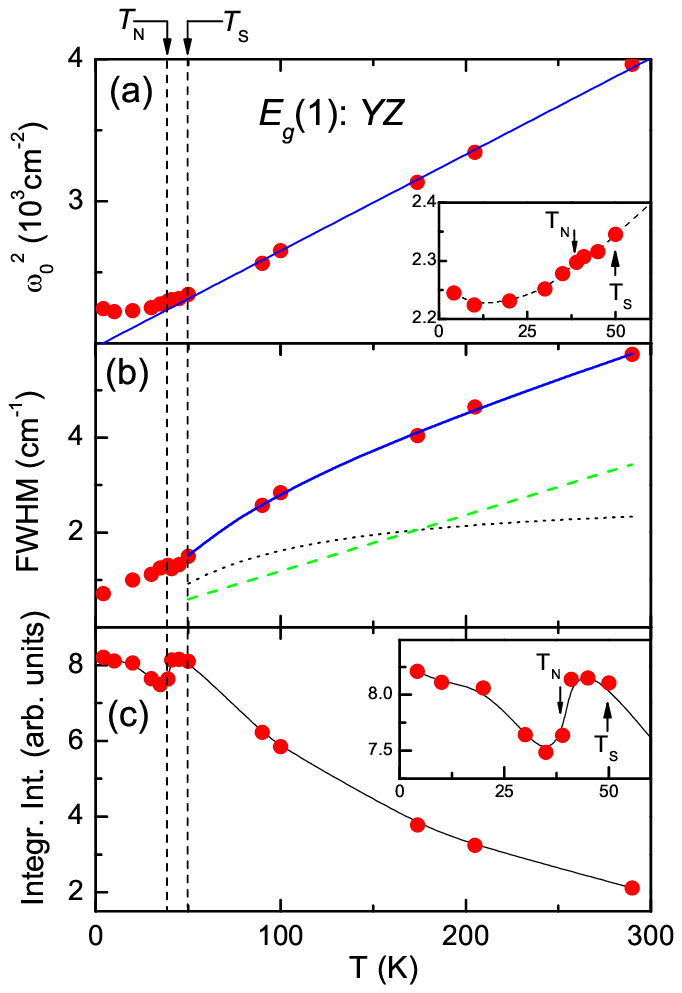}
\caption{\label{fig:AEg1YZoverview} Parameters of the $E_{g}(1)$
mode in the $yz$ scattering geometry: Temperature dependence of (a)
the squared eigenfrequency $\omega_0^2$ together with a fit using
Eq.~\ref{Eq:softmode}, (b) the FWHM linewidth with a fit using
Eq.~\ref{Eq:linewidth} -- dashed and dotted lines: first and second
terms in Eq.~\ref{Eq:linewidth}, respectively, and (c) the Bose
corrected integrated intensity (solid line is a guide to the eye). The
insets highlight the data for $T \leq $60 K.}
\end{figure}

\begin{figure}
 \centering
\includegraphics[width=8cm]{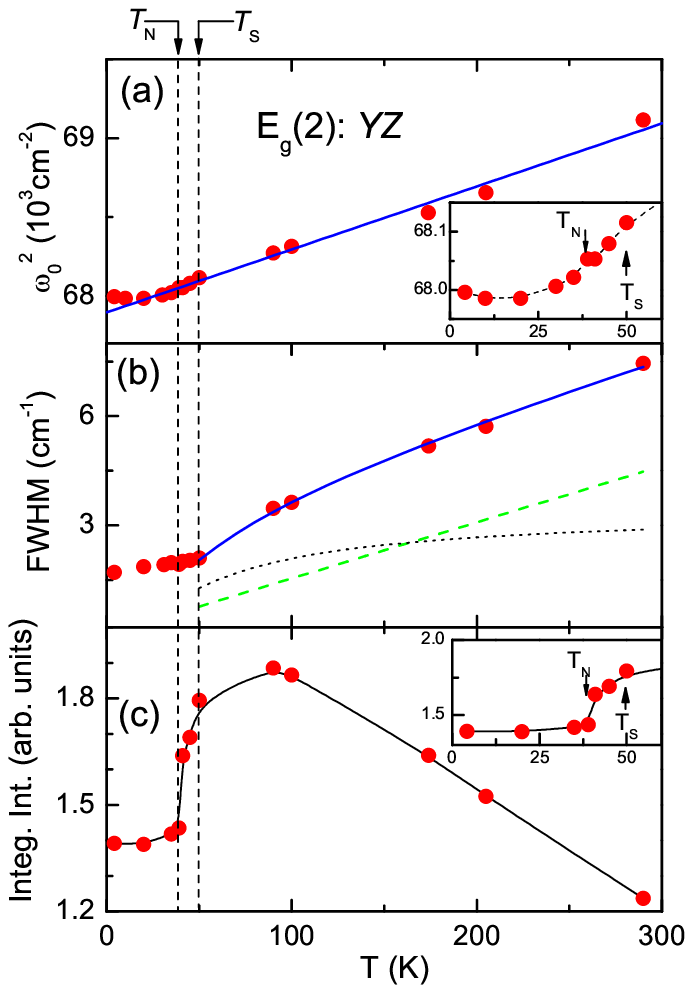}
\caption{\label{fig:Eg2YZoverview} Parameters of the $E_{g}(2)$ mode
in the $yz$ scattering geometry: Temperature dependence of (a) the
squared eigenfrequency $\omega_0^2$ together with a fit using
Eq.~\ref{Eq:softmode}, (b) the FWHM linewidth with a fit using
Eq.~\ref{Eq:linewidth} -- dashed and dotted lines: first and second
terms in Eq.~\ref{Eq:linewidth}, respectively, and (c) the Bose
corrected integrated intensity (line is a guide to the eyes). The
insets highlight the data for $T \leq $60 K.}
\end{figure}


\begin{figure}[h]
 \centering
\includegraphics[width=8cm]{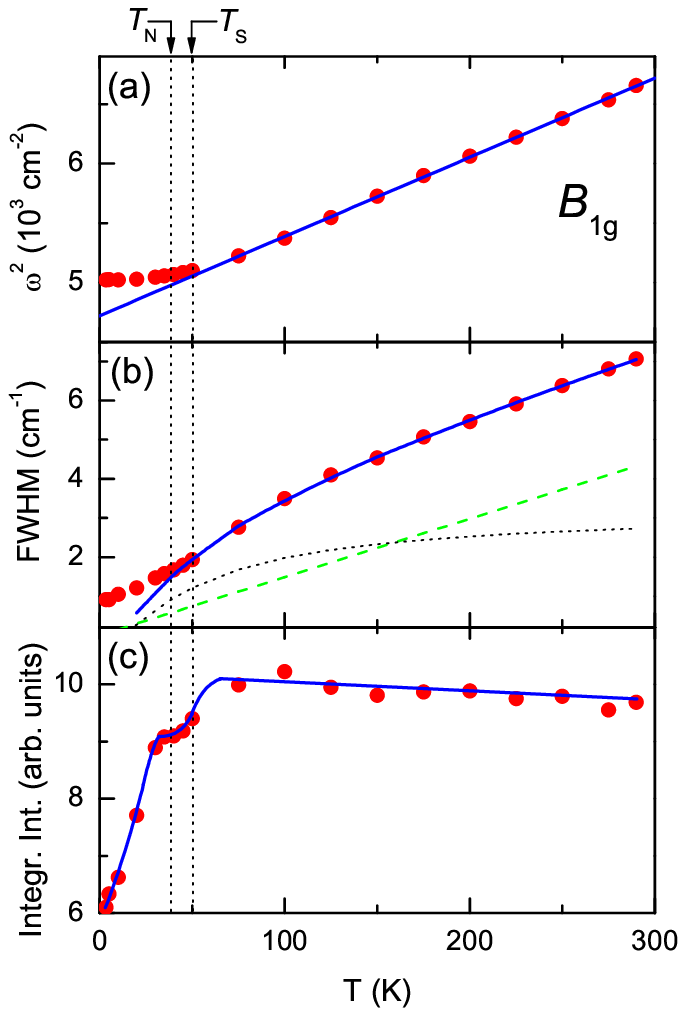}
\caption{\label{fig:B1goverview} Parameters of the $B_{1g}$ mode:
Temperature dependence of (a) the squared eigenfrequency
$\omega_0^2$ together with a fit using Eq.~\ref{Eq:softmode}, (b)
the FWHM linewidth with a fit using Eq.~\ref{Eq:linewidth} -- dashed
and dotted lines: first and second terms in Eq.~\ref{Eq:linewidth},
respectively,, and (c) the Bose corrected integrated intensity and a
solid line to guide the eye.}
\end{figure}

Indeed, when looking at the Raman data of the $E_g$(1) and $E_g$(2)
vibrational modes in $yz$ configuration shown in
Fig.~\ref{fig:fig3}, an anomalous softening of both modes is
observed for $T >$ 50~K. While the frequency shift of the $E_g$(2)
is only about 2 cm$^{-1}$, the low-energy $E_g$(1) mode exhibits a
frequency shift of about 16 cm$^{-1}$. This corresponds to a
softening of 1\% and 25\% with respect to the room temperature
eigenfrequency.

Plotting the square of the eigenfrequency $\omega^2_{E_g(1)}$  of
the $E_g$(1) phonon mode as a function of temperature in
Fig.~\ref{fig:AEg1YZoverview}(a) reveals a linear behavior for $T >$
50~K, which can be understood in terms of a soft-mode behavior
indicative of a structural phase transition expected at $T_\text{c}$
where one expects\cite{Scott74,samara}
\begin{equation}\label{Eq:softmode}
\omega^2_{E_g} =\alpha (T-T_\text{c}).
\end{equation}
The fit shown in Fig.~\ref{fig:AEg1YZoverview}(a) yields  $\alpha$ =
6.8 cm$^{-2}$/K and a virtual transition temperature of $T_\text{c}$
= -291~K. Although the negative sign indicates that the occurrence
of the structural phase transition is very unlikely, the energy
scale of this virtual transition temperature is close to the
orbital-ordering transition temperature $T_{OO}\sim$ 350~K
calculated by assuming a purely electronic superexchange
mechanism.\cite{pavarini}

We believe that the softening of the $E_g$(1) phonon mode is due to
the dynamic nature in the displacement of the apical fluorine ions
away from the $c$ axis, which manifests itself in an anomalously
large thermal displacements parameter\cite{buttner} and the
occurrence of a dynamical Dzyaloshinsky-Moriya (DM)
interaction.\cite{eremin} As a prerequisite for the latter, the
characteristic time of the dynamic distortions must be large
compared to the time scale of the exchange interaction and the
amplitude of these distortions must be high.\cite{eremin} This is
the case for low-lying optical modes contributing to the oscillation
of the F$^-$ ions with the tendency to soften to low temperatures,
exactly like the $E_g$(1) mode. In this scenario the displacement of
the fluorine ions freezes with decreasing temperature and becomes
static at $T_S$= 50~K.\cite{eremin,deisenhofer} Below 50~K we
observe a deviation from this softening behavior and the frequency
levels off in the magnetically ordered state.

The temperature dependence of the phonon line widths full width at
half maximum (FWHM) for $E_g$(1) is shown in
Fig.~\ref{fig:AEg1YZoverview}(b). Above 50~K the linewidth data can
be described (solid line) using:
\begin{equation}\label{Eq:linewidth}
\Gamma_{tot}(T) = \Gamma_{anh}(T) + \Gamma_r(T) = AT + B
\exp\left(-\frac{U_r}{k_BT}\right)
\end{equation}
where $\Gamma_{anh}$ is the contribution arising from phonon
anharmonic interactions in crystalline solids, with zone center
modes decaying into pairs of phonons with equal and opposite wave
vectors.  $\Gamma_r$ is the contribution to the total linewidth
arising from the dynamic deviation of the F$^-$ ions away from the
$c$ axis, $U_r$ is a potential barrier, and $A$ and $B$ are
constants. The data can be described very well over the temperature
range 50 -- 290 K by Eq.~\ref{Eq:linewidth} yielding a energy $U_r$
= 56~K very close to the temperature $T_S$ = 50 K where the dynamic
displacements are proposed to become static.\cite{deisenhofer} Below
$T_S$ = 50 K  the width of the phonon line decrease nearly linearly
with temperature.

The (Bose corrected) integrated intensity of the $E_g$(1) mode shown
in Fig.~\ref{fig:AEg1YZoverview}(c) increases with decreasing
temperature and reaches a maximum at $T_S$ (see inset of
Fig.~\ref{fig:AEg1YZoverview}(c)) and a minimum just below $T_N$
reflecting distinct changes of the polarizability of this mode at
these temperatures.

The corresponding parameters for the $E_g$(2) and the $B_{1g}$ mode
are plotted in Fig.~\ref{fig:Eg2YZoverview} and
Fig.~\ref{fig:B1goverview}, respectively. Similar to the $E_g$(1)
mode these modes exhibit a soft mode behavior with
$\alpha$=3.99~cm$^{-2}$/K and $T_\text{c}$=-17016~K for the $E_g$(2)
and $\alpha$=6.57~cm$^{-2}$/K and $T_\text{c}$=-722~K for the
$B_g$(1) mode. While for the $B_g$(1) with a softening of about 13\%
with respect to room temperature the virtual transition temperature
is still reasonable, the value for the $E_g$(2) mode appears not to
be of physically meaningful due to the moderate softening of only
1-2\%. Note that in other fluorides with rutile structure like
MnF$_2$, NiF$_2$, and FeF$_2$ virtual transition temperatures of
-1240~K, -1700~K, and -1780~K have been derived from the softening
of Raman modes, respectively.\cite{lockwood1,lockwood2,lockwood3}
The linewidth of both modes can again be described using
Eq.\ref{Eq:linewidth} and $U_r$ = 49~K. The intensities of both
modes start to decrease below 50~K, but for the $E_g$(2) mode the
intensity levels off and becomes almost constant below $T_N$.

\begin{figure}[t]
 \centering
\includegraphics[width=8cm]{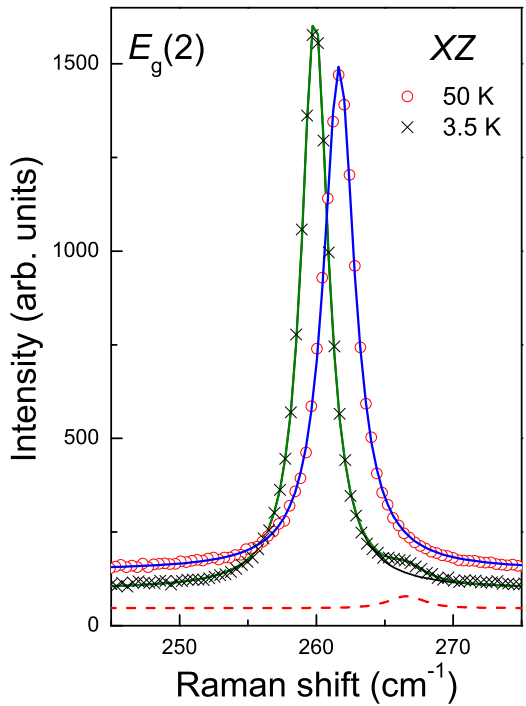}
\caption{\label{fig:Eg2_XZ_spectrum} Raman spectrum of the $E_g$(2)
in $xz$ scattering geometry at 3.5~K (crosses). The solid line and the dashed line (shifted) are the result of fitting the data with two Lorentzian lines. For reference the data at 50~K (open circles) is shown with a single Lorentzian line. }
\end{figure}

\begin{figure}[t]
 \centering
\includegraphics[width=8cm]{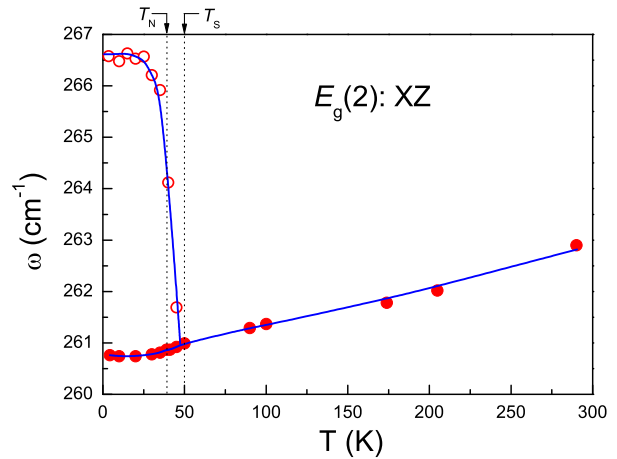}
\caption{\label{fig:Eg2_XZ_freq2} Temperature dependence of the
eigenfrequency of the $E_g$(2) and the split modes below $T_S$ in
$xz$ scattering geometry. Lines are drawn to guide the eye.}
\end{figure}

Although clear anomalies of these modes associated with $T_S$ and
$T_N$ have been observed, we could not observe the splitting of the
$E_g$(1) mode in $yz$-configuration reported in
Ref.~\onlinecite{ueda} at 10~K. Hence, we tried to reproduce the
reported splitting of the $E_g$(2) mode in $xz$-configuration and
trace its temperature dependence. The obtained spectrum at 3.5~K is
shown in Fig.~\ref{fig:Eg2_XZ_spectrum} and a weak additional mode at
the high-frequency side of the $E_g$(2) is clearly visible compared to the data at T=50~K in the same figure. The
solid line corresponds to a fit with two Lorentzian lineshapes
(thin and dashed, latter shifted). As shown in
Fig.~\ref{fig:Eg2_XZ_freq2}, the appearance of this additional mode
coincides with $T_S$ suggesting a splitting of the $E_g$(2) mode in
agreement with the scenario of a symmetry reduction at $T_S$
suggested previously.\cite{deisenhofer} Nevertheless, we have to
point out that this splitting of about 6~cm$^{-1}$ is larger than
the reported one of about 1~cm$^{-1}$ and appears on the high-energy
flank in contrast to the one reported by Ueda and
coworkers,\cite{ueda} which appears on the low-energy side of the
original $E_g$(2) mode. These discrepancies can not be easily
explained and may be due to the different samples used for our work.
We would like to point out that in a recent Raman study a splitting
of the $E_g$(2) similar to our data has been reported.\cite{lee}

A static displacement of fluorine ions away from the $c$ axis at
temperatures $T < T_S$ assumes the lowering of the KCuF$_3$ crystal
symmetry. If the symmetry is lower than $D_{4h}^{18}$, a removal of
the $E_g$ modes' degeneracy and the appearance of extra lines in the
Raman spectra is expected. The observed splitting of the $E_g$(2)
mode in $xz$ configuration confirms this scenario, alone, this
information is not sufficient to determine the low-temperature
symmetry. Additional evidence has been obtained by X-ray scattering
where a splitting of a Bragg reflection associated with GdFeO$_3$
type distortions has been found below 50~K.\cite{lee} Lee and
coworkers also suggested that the observed softening of $E_g$ and
the $B_{1g}$ modes is related to the finite spin correlation lengths
which are inherent to low-dimensional magnets.\cite{lee} Such
effects of spin-phonon coupling are well established and occur,
e.g., in frustrated magnetic systems without orbital degrees of
freedom.\cite{Kant09}

In this respect it is worth highlighting that in KCuF$_3$ the
spin-spin relaxation time as measured by the ESR linewidth can only
be explained by assuming dynamic lattice distortion of the type
associated with the anomalous Raman modes.\cite{eremin} It was also
reported in Ref.~\onlinecite{eremin} that the temperature dependence
of the ESR linewidth $\Delta H$ can be described by  $\Delta
H\propto \exp{-\Delta/T}$ with an activation energy $\Delta$=114~K
which corresponds approximately to $2U_r\approx 2T_S$, two times the
potential barrier derived from the temperature dependence of the
linewidths of the anomalous Raman modes. This intricate feedback
between spin, lattice, and possibly the orbital degrees of freedom
has to be disentangled and the following questions arise and still
need to be clarified: (i) Is the softening of the Raman modes
directly related to the spin-spin correlations of the quasi-one
dimensional spin chain KCuF$_3$? (ii) Is there a relation between
the ESR spin-spin relaxation time dominated by the dynamical
Dzyaloshinsky-Moriya interaction and the linewidth of the anomalous
Raman modes? (iii) How and on which time scale do the orbital
degrees of freedom couple to the lattice and spin fluctuations in
the system? We hope that our study will stimulate further
theoretical efforts in this direction.

\section{Summary}

To sum up, temperature-dependent Raman spectra of single crystalline KCuF$_3$ show a strong softening of the lowest-lying $E_g$(1) and the $B_{1g}$ mode for $T>T_S$. Both of these modes and the $E_g$(2) mode (at about 260~cm$^{-1}$) exhibit anomalies at the characteristic temperature $T_S$=50~K. In $xz$ scattering configuration the $E_g$(2) doublet clearly splits with a splitting of about 6~cm$^{-1}$. The temperature dependence of the linewidth of these modes yields an activated behavior with an energy $U_r\approx$ 50~K corresponding to $T_S$. We ascribe this anomalous behavior and the observed splitting to an antiferrodistortive lattice instability due to strong dynamic displacements of the F$^-$ ions away from the Cu-F-Cu bonding line along the $c$ axis. These displacements are strongly influencing the spin-spin relaxation by allowing for a dynamical Dzyaloshinsky-Moriya interaction. They become static for T$<$$T_S$.

\begin{acknowledgments}
We like to thank M. V. Eremin and B. Lake for useful discussions. V.G. and O.A. acknowledge the Russian-Ukrainian Grant 2009-9 for partial support. D.W. acknowledges support by B-IGSM. We also acknowledge support by the DFG via TRR80 and via LE 967/6-1 and the Swiss NSF through NCCR MaNEP.
\end{acknowledgments}

\end{document}